\documentclass[fleqn,12pt,twoside]{article}
\usepackage[headings]{espcrc1}
\usepackage{amsmath}

\readRCS
$Id: espcrc1.tex,v 1.2 2004/02/24 11:22:11 spepping Exp $
\usepackage{graphicx}

\usepackage[figuresright]{rotating}

\newcommand{\tr}{\hbox{tr}}
\newcommand{\Tr}{\hbox{Tr}}

\def\slashchar#1{\setbox0=\hbox{$#1$}           % set a box for #1 
   \dimen0=\wd0                                 % and get its size
   \setbox1=\hbox{/} \dimen1=\wd1               % get size of /
   \ifdim\dimen0>\dimen1                        % #1 is bigger
      \rlap{\hbox to \dimen0{\hfil/\hfil}}      % so center / in box
      #1                                        % and print #1
   \else                                        % / is bigger
      \rlap{\hbox to \dimen1{\hfil$#1$\hfil}}   % so center #1
      /                                         % and print /
   \fi}

\title{2PI renormalized effective action for gauge theories}

\author{U. Reinosa\address{Institute f\"ur Theoretische Physik, Universit\"at Heidelberg, \\ 
        Philosophenweg 16, 69120 Heidelberg, Germany}%
        }
       
\runtitle{2PI effective ation for gauge theories: Renormalization}
\runauthor{U. Reinosa}

\begin{document}

% typeset front matter
\maketitle

\begin{abstract}
We show how to perform renormalization in the framework of the 2PI effective action for abelian gauge theories. In
addition to the usual renormalization conditions one needs to incorporate new ones in order to remove non-transverse UV
divergences in the truncated two- and four-photon functions. The corresponding counterterms are allowed by gauge
symmetry, in-medium independent and suppressed with respect to the accuracy of the truncation.
\end{abstract}

\section{Introduction}
Functional techniques based on the two-particle-irreducible (2PI) effective action~\cite{Luttinger:1960ua} provide a powerful tool
to devise systematic non-perturbative approximations, of particular interest in numerous physical situations
\cite{Blaizot:2003tw} where
standard expansion schemes are badly convergent \cite{Berges:2004hn}. However, the systematic implementation of 2PI techniques for gauge
theories has been postponed for a long time due to formal intricacies~\cite{Arrizabalaga:2002hn}. Here, we discuss the
issue of renormalization\footnote{For a detailed analysis of this issue in scalar theories see \cite{vanHees:2001ik}.} and
show how to remove UV divergences for any loop approximation of the 2PI effective action. For the sake of simplicity, we
consider QED in the vacuum although our approach could in principle be used for non-abelian gauge theories at finite
temperature or finite chemical potential. We choose a covariant gauge, for which the classical action reads
\begin{equation}
S=\int_x \left\{\bar\psi\Big[i\slashchar{\partial}-e\slashchar{A}-m\Big]\psi+\frac{1}{2}A_\mu\Big[g^{\mu\nu}\partial^2-(1-\lambda)\partial^\mu\partial^{\nu}\Big]A_\nu\right\}\,,
\end{equation}
with $\lambda$ the gauge fixing parameter. A detailed version of this work is found in~\cite{Reinosa:2006cm}.

\section{Two- and four-point functions}
The 2PI effective action for QED in the absence of mean fields is given by
\begin{equation}\label{eq:2PI}
\Gamma_{\rm 2PI}[D,G]=-i\Tr\ln D^{-1}-i\Tr\, D_0^{-1}D+\frac{i}{2}\Tr\ln G^{-1}+\frac{i}{2}\Tr\,G_0^{-1}G+\Gamma_{\rm int}[D,G]\,,
\end{equation}
where the trace {\rm Tr} includes integrals in configuration space and sums over Dirac and Lorentz indices. The free inverse propagators are given by
\begin{eqnarray}
iD_{0,\alpha\beta}^{-1}(x,y) & = & \left[\,i\slashchar{\partial}_x-m\,\right]_{\alpha\beta}\delta^{(4)}(x-y)\,,\\
iG_{0,\mu\nu}^{-1}(x,y) & = & \left[\,g_{\mu\nu}\partial_x^2-(1-\lambda)\partial^x_\mu\partial^x_\nu\,\right]\delta^{(4)}(x-y)\,.
\end{eqnarray}
The functional $\Gamma_{\rm int}[D,G]$ is the infinite series of closed two-particle-irreducible (2PI) diagrams with
lines corresponding to arbitrary two-point functions $D$ and $G$ and with the usual QED vertex. The physical two-point functions $\bar D$ and $\bar G$ can be obtained from the condition that the 2PI functional be stationary, which can be written as
\begin{eqnarray}
 \bar D_{\,\alpha\beta}^{-1}(x,y)-D^{-1}_{0,\alpha\beta}(x,y) & = &
 i\left.\frac{\delta\Gamma_{\rm int}[D,G]}{\delta D^{\beta\alpha}(y,x)}\right|_{\bar D,\bar G}\equiv
 \bar\Sigma_{\alpha\beta}(x,y)\,,\\
 \bar G_{\mu\nu}^{-1}(x,y)- G_{0,\mu\nu}^{-1}(x,y) & = &
 -2i\left.\frac{\delta\Gamma_{\rm int}[D,G]}{\delta G^{\nu\mu}(y,x)}\right|_{\bar D,\bar G}\equiv \bar\Pi_{\mu\nu}(x,y)\,.
\end{eqnarray}
In the next sections we explain how to renormalize $\bar D$ and $\bar G$. In this procedure, an important role is played
by the four-point function with four photon legs which is constructed as follows (see also~\cite{Reinosa:2006cm}). First, one defines the 2PI four-point kernels $\bar\Lambda_{GG}$, $\bar\Lambda_{GD}$, $\bar\Lambda_{DG}$ and $\bar\Lambda_{DD}$ by
\begin{eqnarray}
\bar\Lambda_{GG}^{\mu\nu,\rho\sigma}(p,k) & \equiv &
\left.4\frac{\delta^2\Gamma_{\rm int}[D,G]}
{\delta G_{\nu\mu}(p)\,\delta G_{\rho\sigma}(k)}\right|_{\bar D,\bar G}\,,
\\
\bar\Lambda_{GD}^{\mu\nu;\alpha\beta}(p,k) & \equiv &
\left.-2\frac{\delta^2\Gamma_{\rm int}[D,G]}
{\delta G_{\nu\mu}(p)\,\delta D_{\alpha\beta}(k)}\right|_{\bar D,\bar G}\,,
\\
\bar\Lambda_{DG}^{\alpha\beta;\mu\nu}(p,k) & \equiv &
\left.-2\frac{\delta^2\Gamma_{\rm int}[D,G]}
{\delta D_{\beta\alpha}(p) \,\delta G_{\mu\nu}(k)}\right|_{\bar D,\bar G}\,,
\\
\bar\Lambda_{DD}^{\alpha\beta,\delta\gamma}(p,k) & \equiv &
\left.\frac{\delta^2\Gamma_{\rm int}[D,G]}
{\delta D_{\beta\alpha}(p)\,\delta D_{\delta\gamma}(k)}\right|_{\bar D,\bar G}\,.
\end{eqnarray}
These are then combined in order to build the kernel
\begin{eqnarray}
&&\bar K^{\mu\nu,\rho\sigma}(p,k) =
\bar \Lambda_{GG}^{\mu\nu,\rho\sigma}(p,k)
-\int_q \bar \Lambda_{GD}^{\mu\nu;\alpha\beta}(p,q)M^{DD}_{\alpha\beta,\gamma\delta}(q)
\bar \Lambda_{DG}^{\gamma\delta;\rho\sigma}(q,k)\nonumber\\
&&\qquad+
\int_q\int_r\bar\Lambda_{GD}^{\mu\nu;\alpha\beta}(p,q)M^{DD}_{\alpha\beta,\bar\alpha\bar\beta}(q)
\bar \Lambda^{\bar\alpha\bar\beta,\bar\gamma\bar\delta}(q,r)
M^{DD}_{\bar\gamma\bar\delta,\gamma\delta}(r)\bar \Lambda_{DG}^{\gamma\delta;\rho\sigma}(r,k)\,,
\end{eqnarray}
where $M^{DD}_{\alpha\beta,\gamma\delta}(q)\equiv\bar D_{\alpha\gamma}(q)\bar D_{\delta\beta}(q)$ and
\begin{equation}
 \bar \Lambda^{\alpha\beta,\gamma\delta}(p,k)=
 \bar \Lambda_{DD}^{\alpha\beta,\gamma\delta}(p,k)
 -\int_q\bar\Lambda_{DD}^{\alpha\beta,\bar\alpha\bar\beta}(p,q)
 M^{DD}_{\bar\alpha\bar\beta,\bar\gamma\bar\delta}(q)
 \bar\Lambda^{\bar\gamma\bar\delta,\gamma\delta}(q,k)\,.
\end{equation}
The kernel $\bar K$ is 2PI
with respect to photon lines but 2PR (two-particle-reducible) with respect to fermion lines. Finally the four-photon
function $\bar V$ is obtained from the equation
\begin{equation}
\bar V^{\mu\nu,\rho\sigma}(p,k)=\bar K^{\mu\nu,\rho\sigma}(p,k)+\frac{1}{2}\int_q
\bar K^{\mu\nu,\bar\mu\bar\nu}(p,q)M^{GG}_{\bar\mu\bar\nu,\bar\rho\bar\sigma}(q)
\bar V^{\bar\rho\bar\sigma,\rho\sigma}(q,k)\,,
\end{equation}
where $M^{GG}_{\mu\nu,\rho\sigma}(q)=\bar G_{\mu\rho}(q)\bar G_{\sigma\nu}(q)$. The four-photon function $\bar V$ is 2PR both with respect to fermion and photon lines.

\section{Gauge symmetry and counterterms}
In the absence of mean fields, the 2PI effective action satisfies the Ward-Takahashi identity $\Gamma_{\rm int}[G^\alpha,D^\alpha]=\Gamma_{\rm int}[G,D]$ with $G^\alpha(x,y) \equiv G(x,y)$ and $D^\alpha(x,y) \equiv e^{i\alpha(x)}D(x,y) e^{-i\alpha(y)}$\,. In view of eliminating divergences in $\bar D$ and $\bar G$, one modifies the 2PI effective action by adding a shift $\delta \Gamma_{\rm int}[D,G]$ compatible with all the symmetries of the system, namely Lorentz and gauge symmetry. At two-loop order the shift reads for instance:
\begin{eqnarray}
 \delta\Gamma_{\rm int}[D,G]&=&\int_x\left\{
 -\tr\Big[(i\delta Z_2\slashchar{\partial}_x -\delta m)D(x,y)\Big]
 +\frac{\delta Z_3}{2}\,\Big(g^{\mu\nu}\partial_x^2
 -\partial_x^\mu\partial_x^\nu\Big)G_{\mu\nu}(x,y)\right.\nonumber\\
 &&\qquad+\frac{\delta\lambda}{2}\,\partial_x^\mu\partial_x^\nu G_{\mu\nu}(x,y)
 +\frac{\delta M^2}{2}\,{G^\mu}_\mu(x,x)\nonumber\\
 &&\qquad\left.+\,\frac{\delta g_1}{8}\,{G^\mu}_\mu(x,x){G^\nu}_\nu(x,x)+
 \frac{\delta g_2}{4}\,G^{\mu\nu}(x,x)G_{\mu\nu}(x,x)\right\}_{\!\!y=x}\,,
\end{eqnarray}
where ${\rm tr}$ denotes the trace over Dirac indices. The counterterms $\delta Z_2$, $\delta Z_3$ and $\delta m$ are
the analog of the corresponding  ones in perturbation theory.\footnote{Notice that the counterterm $\delta e$ only
appears at higher orders in the 2PI loop-expansion.} The extra counterterms $\delta\lambda$,
$\delta M^2$, $\delta g_1$ and $\delta g_2$ have no analog in perturbation theory but are allowed by the 2PI Ward-Takahashi
identity. Their role is to absorb non-transverse divergences in the two- and four-photon functions. At higher
loops, additional diagrams including counterterms need to be considered.

\section{Renormalization conditions}
If we were to work with the exact 2PI effective action (no truncations), it would be easy to check that the two-photon
and four-photon functions $\bar \Pi$ and $\bar V$ are transverse. This would prevent the appearance of non-transverse divergences in $\bar \Pi$ and
$\bar V$ and would lead to $\delta\lambda=\delta M^2=\delta g_1=\delta g_2=0$, as one is used to in perturbation theory.
However, one has in general to approximate the 2PI effective action by truncating its diagrammatic expansion to a certain
loop order. This leads to non-transverse contributions in the two- and four-photon functions. For instance, at two-loop
order order, one can check that divergent non-transverse contributions appear in $\bar V$ at order $e^4$:
\begin{equation}
\bar V_{\rm 2-loop}^{\rm \mu\nu,\,\rho\sigma,\,
div}=\frac{e^4}{\pi^2}\frac{d-2}{d+2}\frac{g^{\mu\rho}g^{\nu\sigma}+g^{\mu\sigma}g^{\nu\rho}}{d-4}\,,
\end{equation}
where $d$ represents the dimension of space in dimensional regularization. In general, at $L$-loop order,
non-transverse divergences appear at order $e^{2L}$ and one has to devise a procedure to consistently remove them.

First one needs to remove four-photon sub-divergences in $\bar D$ and $\bar G$. A diagrammatic analysis reveals that
these are nothing but the ones encoded in $\bar V$.\footnote{The connection between $\bar D$, $\bar G$ and $\bar V$ is
also easily seen at finite temperature where requiring that $\bar D$ and $\bar G$ do not contain temperature dependent divergences is equivalent to requiring that $\bar V$ is finite.} Renormalization of $\bar V$ is then achieved by tuning $\delta g_1$ and $\delta g_2$ via the renormalization condition
\begin{equation}
 P_{L,\mu\nu}(k_*)\bar V^{\mu\nu,\rho\sigma}(k_*,k_*)=0\,,
\end{equation}
where $P_L^{\mu\nu}(k_*)=n_*^\mu\,n_*^\nu$ is the longitudinal projector, with $n_*^\mu\equiv
k_*^\mu/\sqrt{k_*^2}$. Notice that, as the number of loops goes to infinity, the renormalization condition becomes an identity (1PI
Ward-Takahashi identity), ensuring that no new parameter is introduced and that $\delta g_1 \rightarrow 0$ and $\delta
g_2 \rightarrow 0$, as they should.

Once sub-divergences have been eliminated, there only remain overall divergences in $\bar D$ and $\bar G$.\footnote{At
finite temperature or finite chemical potential, these divergences are in-medium independent. This is the benefit of
properly removing four-photon sub-divergences first.} Divergences in
$\bar D$ are removed as usual via $\delta m$ and $\delta Z_2$ which are fixed by imposing the renormalization conditions $\bar{\Sigma}(p=p_\star)=0$ and $d\bar{\Sigma}(p)/d\slashchar{p}|_{p=p_\star}=0$. Similarly,
for the photon self-energy, one can impose the following condition on the transverse part of the photon polarization
tensor $d\bar{\Pi}_{\rm T}(k^2)/dk^2|_{k=k_*}=0$.
Usually (in perturbation theory for instance), this is enough to fix the photon wave-function renormalization counterterm and to remove all divergences, which are purely transverse. In the 2PI
framework, however, non-transverse divergences appear and one needs to fix three counterterms $\delta Z_3$, $\delta\lambda$ and $\delta M^2$. Two other
independent conditions are thus needed. Again, we choose renormalization conditions which are automatically fulfilled in the
exact theory in order to ensure that $\delta\lambda \rightarrow 0$ and $\delta M^2 \rightarrow 0$, in the limit of
an infinite number of loops. A possible choice is to
impose transversality of the photon polarization tensor:
\begin{equation}
 \bar{\Pi}_{\rm L}(k^2_*)=0\,.
\end{equation}
Since in the exact theory the transversality condition has to hold for all momenta, one can further impose that
\begin{equation}
 \left.\frac{d\bar\Pi_{\rm L}(k^2)}{d k^2}\right|_{k=k_*}=0\,.
\end{equation}

\end{document}